# How Much Rate Splitting Is Required for a Random Coding Scheme?

## A New Achievable Rate Region for the Broadcast Channel with Cognitive Relays

Reza K. Farsani

School of Cognitive Sciences, Institute for Research in Fundamental Sciences (IPM), Tehran, Iran

Email: reza_khosravi@alum.sharif.ir

*Abstract*—In this paper, it is shown that for any given single-hop communication network with two receivers, splitting messages into more than two sub-messages in a random coding scheme is redundant. To this end, the Broadcast Channel with Cognitive Relays (BCCR) is considered. A novel achievability scheme is designed for this network. Our achievability design is derived by a systematic combination of the best known achievability schemes for the basic building blocks included in the network: the Han-Kobayashi scheme for the two-user interference channel and the Marton's coding scheme for the broadcast channel. Meanwhile, in our scheme each private message is split into only two sub-messages which is identically exploited also in the Han-Kobayashi scheme. It is shown that the resultant achievable rate region includes previous results as well. More importantly, the procedure of the achievability design is described by graphical illustrations based on directed graphs. Then, it is argued that by extending the proposed scheme on the MACCM plan of messages, one can derive similar achievability schemes for any other single-hop communication network.

## I. INTRODUCTION

One of the basic problems in network information theory is to derive achievability schemes for communication networks. This problem has been widely explored for simple network topologies, however, there is a very few results for large multi-message scenarios. Given a network topology of arbitrary dimensions, an essential question is how to design an achievability scheme with a satisfactory performance. For large networks, there exist numerous methods to build coding schemes. But what is really the best strategy? To illustrate the importance of the problem, let us discuss an example. The cognitive radio channel was introduced in 2006 in [10] where the authors established an achievable rate region for the channel by a combination of the Gel'fand-Pinsker binning technique and the Han-Kobayashi rate splitting scheme. Following up this paper, many other researchers [11-22] studied information theoretic bounds for this channel. Specifically, several achievability schemes were proposed for the channel in these papers. But it was not clear which of the proposed achievable rate regions is better than the others because their comparisons seem to be difficult. Recently in [13], the authors presented a new achievability scheme and showed that it includes the previous results. Even for the case of [13], due to the lack of systematic derivation, some extra complexities have been introduced in the achievability scheme. In the proposed scheme, the primary transmitter splits its message into *three parts* and then one of these sub-messages is ignored and relegated to the cognitive transmitter. In [1, Sec. III.A.4], we have demonstrated that this strategy, i.e., splitting the primary message into three parts, not only is superfluous, but also it may cause rate loss. In fact, the authors in [13] have derived their achievability scheme by a combination of previous ones. As the cognitive radio channel is rather a simple topology, it is clear that for larger networks the problem is more intricate. Therefore, it is essential to have suitable criteria to build achievability schemes. We provide a novel framework to systematically design achievability schemes based on the following criteria:

✓ Given a certain network, our achievability design is such that when it is specialized for the basic building blocks contained in the network, it does work essentially similar to the best known coding strategies.

✓ To reduce the complexity, to the extent possible without violating the previous criterion, the achievability scheme is designed with the lowest number of message splitting. In the general case, in [3, Sec. VI] and [5], we prove that for an intelligent achievability deign for a network with $K$ receivers, each message is required to be split at most into $2^{K-1}$ parts (each sub-message is designated to be transmitted to the desired receiver as well as a subset of non-desired receivers). Specially, for networks with two receivers, splitting each message into more than two parts is always redundant.

✓ The superposition structures among the generated codewords are such that each receiver is required to decode only its designated codewords. Clearly, when two codewords build a superposition structure, for correct decoding of the satellite codeword it is required to decode the cloud center first. Now if a satellite codeword is designated for some specific receivers but its cloud center is not, then those receivers should necessarily decode some non-desired codewords, which irrevocably causes rate loss.

Based on the above criteria, we design a unique achievability scheme for all single-hop communication networks with arbitrary topologies. This design is derived by a systematic combination of the best achievability schemes for the Multiple Access Channel (MAC) and the Broadcast Channel (BC) with common messages. Moreover, we provide a graphical illustration for our coding strategy using the *MACCM graphs* developed in our paper [2]. Indeed, our graphical approach is a powerful tool not only to describe a given scheme for a certain network but also to design achievability schemes with satisfactory performance for large networks.

In this paper, we intend to address some of the key features of our systematic design for the two-receiver networks. Clearly, we intend to show that for any given single-hop communication network with two receivers, splitting messages into more than two sub-messages in a random coding scheme is redundant. For this purpose, we consider the Broadcast Channel with Cognitive Relays (BCCR). We propose a novel achievability scheme for this network. Our achievability design is derived by a systematic combination of the best known achievability schemes for the basic building blocks included in the network: the Han-Kobayashi scheme for the two-user classical interference channel and the Marton's coding scheme for the broadcast channel. Meanwhile, in our scheme each private message is split into only two sub-messages which is identically exploited also in the Han-Kobayashi scheme. We demonstrate that our achievable rate region includes previous results as well. More importantly, we describe the philosophy behind our achievability design by using the graphical illustrations developed in our previous work [7-8]. Then, we argue that by extending the proposed approach on the *MACCM plan of messages* [2], one can derive similar coding schemes for any other interference network.

In this paper, Random Variables (RV) are denoted by upper case letters (e.g. $X$) and lower case letters are used to show their realization (e.g. $x$). The Probability Distribution Function (PDF) of $X$ is denoted by $P_X(x)$ and the conditional PDF of $X$ given $Y$ is denoted by $P_{X|Y}(x|y)$. Also, information theoretic concepts such as achievable rate region and capacity are defined in the standard Shannon sense [9].

## II. MAIN RESULTS

Consider the BCCR with common message shown in Fig. 1. This interference network is composed of three transmitters and two receivers. The transmitter $X_B$ (broadcasting node) sends three messages $M_0, M_1, M_2$ to the receivers $Y_1$ and $Y_2$ while being assisted by two relay transmitters $X_1$ and $X_2$. The relay $X_1$ has access to the message $M_1$ and the relay $X_2$ to the message $M_2$. The messages $M_0, M_1$ are decoded at the receiver $Y_1$ and the messages $M_0, M_2$ at the receiver $Y_2$. The conditional probability function $\mathbb{P}(y_1, y_2|x_1, x_B, x_2)$ describes the relation between inputs and outputs of the network.

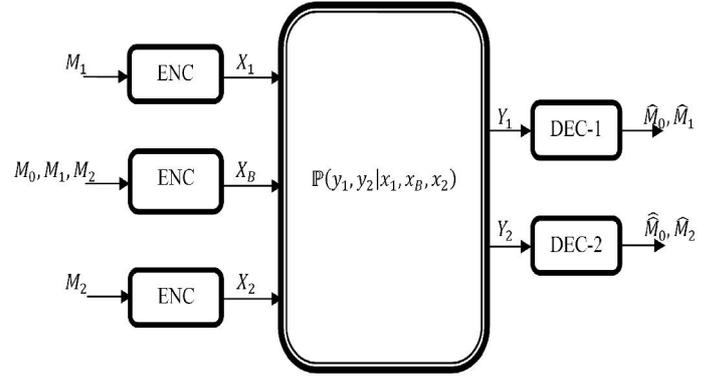

Figure 1. The Broadcast Channel with Cognitive Relays (BCCR).

Note that the BCCR contains the two-user interference channel, the cognitive radio channel [10] and the two-user BC with common message as special cases. Moreover, this model is reduced to the one considered in [19] when there is no common message, i.e., $M_0 \equiv \emptyset$.

It is clear that for such a large network, one can propose numerous achievability schemes based on random coding techniques. But what is the best strategy? In the following theorem, we present our achievability scheme for the BCCR and then deal with about its capabilities. In fact, we intend to present a glimpse of our general random coding scheme in [5].

***Theorem 1)*** *Consider the BCCR with common message in Fig. 1. Define the rate region $\mathfrak{R}_i^{F \to BCCR_{cm}}$ as given in (1), where $\mathcal{P}^{BCCR}$ is the set of all joint PDFs $P_{QW_1U_1W_2V_2W_BU_BV_BX_1X_2X_B}$ which are factorized according to (2). The set $\mathfrak{R}_i^{F \to BCCR_{cm}}$ constitutes an inner bound on the capacity region.*

$$\mathfrak{R}_i^{F \to BCCR_{cm}} \triangleq \bigcup_{\mathcal{P}^{BCCR}} \left\{ \begin{array}{l} (R_0, R_1, R_2) \in \mathbb{R}_+^3 : \exists (B_0, B_1, B_2, R_0, R_{10}, R_{11}, R_{20}, R_{22}) \in \mathbb{R}_+^8 : \\ R_1 = R_{10} + R_{11}, \quad R_2 = R_{20} + R_{22}, \\[4pt] B_0 > I(U_1, V_2; W_B|W_1, W_2, Q) \triangleq \eta_0 \\ B_0 + B_1 > \eta_0 + I(V_2; U_B|U_1, W_1, W_2, W_B, Q) \triangleq \eta_0 + \eta_1 \\ B_0 + B_2 > \eta_0 + I(U_1; V_B|V_2, W_1, W_2, W_B, Q) \triangleq \eta_0 + \eta_2 \\ B_0 + B_1 + B_2 > \eta_0 + \eta_1 + \eta_2 + I(U_B; V_B|U_1, V_2, W_1, W_2, W_B, Q) \\ \theta_1 \triangleq I(U_1; W_B|W_1, W_2, Q), \\ R_{11} + B_1 < I_{Y_1}^1 + \theta_1 \triangleq I(U_1, U_B; Y_1|W_1, W_2, W_B, Q) + \theta_1 \\ R_0 + B_0 + B_1 < I_{Y_1}^2 + \theta_1 \triangleq I(W_B, U_B; Y_1|W_1, W_2, U_1, Q) + \theta_1 \\ R_{20} + R_0 + B_0 + B_1 < I_{Y_1}^3 + \theta_1 \triangleq I(W_2, W_B, U_B; Y_1|W_1, U_1, Q) + \theta_1 \\ R_0 + B_0 + R_{11} + B_1 < I_{Y_1}^4 + \theta_1 \triangleq I(U_1, W_B, U_B; Y_1|W_1, W_2, Q) + \theta_1 \\ R_{20} + R_0 + B_0 + R_{11} + B_1 < I_{Y_1}^5 + \theta_1 \triangleq I(U_1, W_2, W_B, U_B; Y_1|W_1, Q) + \theta_1 \\ R_{10} + R_0 + B_0 + R_{11} + B_1 < I_{Y_1}^6 + \theta_1 \triangleq I(W_1, U_1, W_B, U_B; Y_1|W_2, Q) + \theta_1 \\ R_{10} + R_{20} + R_0 + B_0 + R_{11} + B_1 < I_{Y_1}^7 + \theta_1 \triangleq I(W_1, U_1, W_2, W_B, U_B; Y_1|Q) + \theta_1 \\ \theta_2 \triangleq I(V_2; W_B|W_1, W_2, Q) \\ R_{22} + B_2 < I_{Y_2}^1 + \theta_2 \triangleq I(V_2, V_B; Y_2|W_1, W_2, W_B, Q) + \theta_2 \\ R_0 + B_0 + B_2 < I_{Y_2}^2 + \theta_2 \triangleq I(W_B, V_B; Y_2|W_1, W_2, V_2, Q) + \theta_2 \\ R_{10} + R_0 + B_0 + B_2 < I_{Y_2}^3 + \theta_2 \triangleq I(W_1, W_B, V_B; Y_2|W_2, V_2, Q) + \theta_2 \\ R_0 + B_0 + R_{22} + B_2 < I_{Y_2}^4 + \theta_2 \triangleq I(V_2, W_B, V_B; Y_2|W_1, W_2, Q) + \theta_2 \\ R_{10} + R_0 + B_0 + R_{22} + B_2 < I_{Y_2}^5 + \theta_2 \triangleq I(W_1, V_2, W_B, V_B; Y_2|W_2, Q) + \theta_2 \\ R_{20} + R_0 + B_0 + R_{22} + B_2 < I_{Y_2}^6 + \theta_2 \triangleq I(W_2, V_2, W_B, V_B; Y_2|W_1, Q) + \theta_2 \\ R_{10} + R_{20} + R_0 + B_0 + R_{22} + B_2 < I_{Y_2}^7 + \theta_2 \triangleq I(W_1, V_2, W_2, W_B, V_B; Y_2|Q) + \theta_2 \end{array} \right\}$$

(1)

$$\mathcal{P}^{BCCR} : P_Q P_{W_1 U_1 X_1|Q} P_{W_2 V_2 X_2|Q} P_{W_B U_B V_B|W_1 U_1 W_2 V_2 Q} P_{X_B|X_1 X_2 V_B U_B W_B U_1 W_1 V_2 W_2 Q}$$

(2)

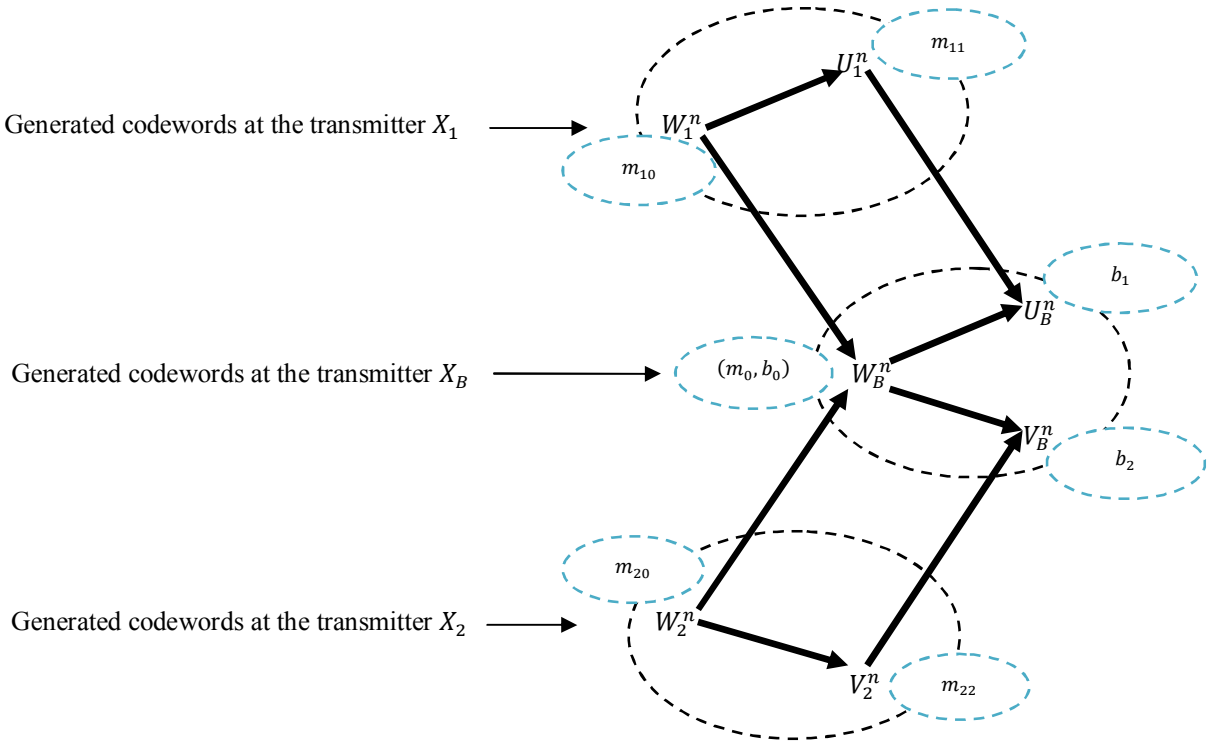

Figure 2. The graphical illustration of the achievability scheme for the BCCR with common message.

*Proof of Theorem 1)* The rate region $\mathfrak{R}_i^{F \to BCCR_{cm}}$ is derived using a random coding strategy. In this paper, we only outline key elements of our achievability scheme. The analysis of the corresponding error probability can be performed directly using the general expressions given in our previous paper [7, Appendix]. This analysis can also be found in [3, Sec. VI].

The achievability scheme includes a rate splitting technique. Each of the messages $M_1$ and $M_2$ and thereby their respective communication rates $R_1$ and $R_2$ are split into only two parts:

$$M_1 = (M_{10}, M_{11}), \qquad R_1 = R_{10} + R_{11}$$
$$M_2 = (M_{20}, M_{22}), \qquad R_2 = R_{20} + R_{22} \qquad (3)$$

Similar to the Han-Kobayashi scheme one part of each private message, i.e., $M_{10}$ of $M_1$ and $M_{20}$ of $M_2$, is used to contribute in building a joint decoding at the non-respective receiver. The split messages are then encoded by a random code of length-$n$. Roughly speaking, in our coding scheme the transmitter $X_1$ encodes its corresponding messages $M_{10}$ and $M_{11}$ by the codewords $W_1^n$ and $U_1^n$, respectively, in a superposition style such that $W_1^n$ serves as the cloud center and $U_1^n$ as the satellite codeword. Similarly, the transmitter $X_2$ (independent of the transmitter $X_1$) encodes its corresponding messages $M_{20}$ and $M_{22}$ by the codewords $W_2^n$ and $V_2^n$, respectively, in a superposition style where $W_2^n$ serves as the cloud center and $V_2^n$ as the satellite. At the transmitter $X_B$, three bins of codewords are generated:

- $2^{nB_0}$ codewords $W_B^n$ conveying the common message $M_0$ which are superimposed upon $W_1^n, W_2^n$.
- $2^{nB_1}$ codewords $U_B^n$ which are superimposed upon $W_1^n, W_2^n, W_B^n, U_1^n$.
- $2^{nB_2}$ codewords $V_B^n$ which are superimposed upon $W_1^n, W_2^n, W_B^n, V_2^n$.

In fact, the codewords $U_B^n$ and $V_B^n$ do not convey any part of the messages beyond than those conveyed by their cloud centers. These are satellite codewords (with the cloud centers $W_1^n, W_2^n, W_B^n, U_1^n$ for $U_B^n$, and $W_1^n, W_2^n, W_B^n, V_2^n$ for $V_B^n$) which are served just for building a Marton's type encoding (refer to our paper presented in [8]) at the transmitter $X_B$ (broadcast node). Fig. 2 represents the generated codewords at the transmitters. Similar to our graphical illustrations in [8], in this figure every two codewords connected by a directed edge build a superposition structure: The codeword at the beginning of the edge is the cloud center and the one at the end of the edge serves as the satellite codeword. The ellipse beside each codeword shows what is conveyed by that codeword in addition to the ones conveyed by its cloud centers. The sizes of the bins $B_0, B_1, B_2$ are selected such large to guarantee the existence of a 7-tuple $(W_1^n, U_1^n, W_2^n, V_2^n, W_B^n, U_B^n, V_B^n)$ of jointly typical (with respect to the distribution (2)) codewords for each given 5-tuple $(M_0, M_{10}, M_{11}, M_{20}, M_{22})$ of messages. These typical codewords are designated for transmission. The transmitter $X_1$ then generates a codeword $X_1^n$ superimposed upon its designated codewords $W_1^n, U_1^n$, and sends over the network. Similarly, the transmitter $X_2$ generates a codeword $X_2^n$ superimposed upon its corresponding codewords $W_2^n, U_2^n$, and transmit it. As the transmitter $X_B$ has access to the codewords of the relay transmitters $X_1$ and $X_2$, it generates a codeword $X_B^n$ superimposed upon all the other designated codewords, i.e., $X_1^n, W_1^n, U_1^n, X_2^n, W_2^n, V_2^n, W_B^n, U_B^n, V_B^n$, and sends over the network. For decoding, each receiver makes use of a jointly typical decoder. The receiver $Y_1$ explores within the codewords $W_1^n, W_2^n, W_B^n, U_1^n, U_B^n$, to find its intended messages, and the receiver $Y_2$ explores within the codewords $W_1^n, W_2^n, W_B^n, V_2^n, V_B^n$. ∎

But what is the philosophy behind this achievability design? We know that the best coding strategies for the two-user interference channel is due to Han and Kobayashi and that for the two-user BC is due to Marton, both of them have been discussed in details in [1-2] (see also [9]). Now it is reasonable that an efficient achievability scheme for the BCCR, which contains both the interference channel and the BC as special cases, should include the benefits of the Han-Kobayashi scheme and the Marton's scheme simultaneously. It is clear that one may split the messages $M_1$ and $M_2$ into numerous sub-messages and then manipulate them in an order to build a combination of the Han-Kobayashi scheme and the Marton's scheme. In fact, such naive techniques are always available to combine coding schemes of simple channels and derive an achievable rate region for a given large network. Nonetheless, clearly this approach is never efficient. It should be noted that splitting messages into sub-messages yields achievable rate regions with complicated descriptions, specifically for large networks. Indeed, in achievability schemes built by random coding techniques as the number of split messages increases, both the number of parameters involved in the resultant achievable rate region and the number of constraints included in it also (rapidly) increases. Accordingly, evaluation of such achievable rate regions is too difficult. Moreover, as we demonstrated in [1, Sec. III.A.4], splitting a message into several sub-message does not necessarily leads to a larger achievable rate region. In some cases it causes also some rate loss. For example, as discussed in introduction, in [13] the authors derived an achievable rate region for the cognitive radio channel wherein the message of the primary transmitter is split into two parts and the message of the cognitive transmitter is split into three parts. The primary transmitter then ignores one of its sub-messages and relegates its transmission to the cognitive transmitter. In [1, Sec. III.A.4], we showed that such type of message splitting is superfluous and by splitting each message just into two parts a larger achievable rate region is derived.

Now let us examine our achievability scheme designed in Theorem 1 for the BCCR. Refer to Fig. 2. In this scheme we have split each message just into two parts, similar to the Han-Kobayashi achievability scheme. According to the description of our achievability scheme given in the proof of Theorem 1, it is clear that by setting $W_B \equiv U_B \equiv V_B \equiv \emptyset$ and also $M_0 \equiv \emptyset$, we obtain the Han-Kobayashi rate region for the two-user interference channel. Now, consider the codewords generated at the broadcasting node, i.e., the transmitter $X_B$. As discussed in the description of the coding scheme, the codewords $U_B^n$ and $V_B^n$ do not contain any part of the messages other than those conveyed by their cloud centers. But how the achievability scheme presented in Fig. 2 includes the Marton's scheme when it is specialized for the two-user BC? Let us in our scheme set $W_1 \equiv U_1 \equiv \emptyset$ and also $W_2 \equiv V_2 \equiv \emptyset$. What happens by this choice? First note that, however the messages $M_{10}, M_{11}, M_{20}, M_{22}$ are conveyed by the codewords $W_1^n, U_1^n, W_2^n, V_2^n$, respectively; but this does not mean that by setting $W_1 \equiv U_1 \equiv W_2 \equiv V_2 \equiv \emptyset$, the messages $M_{10}, M_{11}, M_{20}, M_{22}$ have been withdrawn from the transmission scheme. These messages still are contained by the codewords $W_B^n, U_B^n, V_B^n$. Precisely speaking, when two codewords build a superposition structure then the satellite codeword includes also those messages and bin indices which are conveyed by the cloud center. For example, in Fig. 2, the codeword $V_B^n$ actually conveys the triple $(m_{10}, m_{11}, b_1)$; note that in Fig. 2, the ellipse beside each codeword only shows what is conveyed by that codeword in addition to the ones conveyed by its cloud centers. Now when we set $W_1^n \equiv U_1^n \equiv W_2^n \equiv V_2^n \equiv \emptyset$, the task of the transmission of those messages conveyed by these codewords, i.e., $M_{10}, M_{11}, M_{20}, M_{22}$, automatically is transferred to their respective satellite codewords. In other words, the coding scheme is reduced to the one shown in Fig. 3.

Also, note that when we set $W_1 \equiv U_1 \equiv W_2 \equiv V_2 \equiv \emptyset$ in the achievable region $\Re_i^{F \to BCCR_{cm}}$ given by (1), we obtain $B_0 = 0$; therefore, $b_0$ is not included in the ellipse beside the codeword $W_B^n$ in Fig. 3. Let compare the scheme in Fig. 3 with the Marton's coding for the two-user BC with common message as illustrated in the Fig. 4 (see [8]). Here, we briefly review the Marton's coding scheme. Consider broadcasting the messages $M_0, M_1, M_2$ by a transmitter $X$ to two receivers $Y_1$ and $Y_2$ where the receiver $Y_1$ is required to decode the messages $M_0, M_1$ and the receiver $Y_2$ to decode the messages $M_0, M_2$. Roughly speaking, in the Marton's coding scheme (for a length-$n$ code) the common message $M_0$ is encoded by a codeword $W^n$ generated based on $P_W$. For each of the private messages, a bin of codewords is randomly generated which are superimposed upon the common message codeword $W^n$: The bin corresponding to $M_1$ contains the codewords $U^n$ generated based on $P_{U|W}$ and that one for $M_2$ contains the codeword $V^n$ generated based on $P_{V|W}$. The sizes of the bins are selected such large to guarantee that for each triple $(M_0, M_1, M_2)$ there exists a triple $(W^n, U^n, V^n)$ jointly typical with respect to the PDF $P_{WUV}$. Superimposed on the designated jointly typical codewords $W^n, U^n, V^n$, the encoder then generates its codewords $X^n$ based on $P_{X|WUV}$, and sends it over the channel. The receiver $Y_1$ decodes the codewords $W^n, U^n$ and the receiver $Y_2$ decodes $W^n, V^n$, both using a jointly typical decoder. At the last step, the resulting achievable rate region is further enlarged by the fact that if the rate triple $(R_0, R_1, R_2) \in \mathbb{R}_+^3$ is achievable for the BC, then $(R_0 - \tau_1 - \tau_2, R_1 + \tau_1, R_2 + \tau_2) \in \mathbb{R}_+^3$, where $\tau_1, \tau_2 \in \mathbb{R}_+$, is also achievable. This latter technique can be re-interpreted as follows: Each of the messages $M_1$ and $M_2$ are split into two parts as in (3) and then the parts $M_{10}$ and $M_{20}$ are transferred to the common message codeword, i.e., $W^n$. In other words, we allow that each receiver decodes a part of its non-respective private message. Therefore, the scheme in Fig. 3 essentially does work similar to the Marton's scheme.

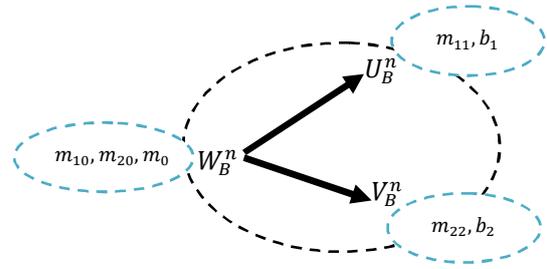

Figure 3. The coding scheme that is derived when we set $W_1^n \equiv U_1^n \equiv W_2^n \equiv V_2^n \equiv \emptyset$ in the scheme of Fig. 2.

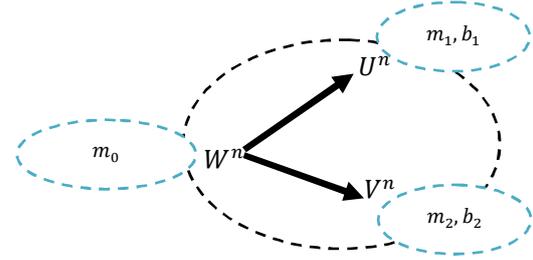

Figure 4. The Marton's coding scheme for the BC with common message.

The above discussion demonstrates the capability of our achievability scheme for the BCCR presented in Theorem 1.

Our design systematically combines the Han-Kobayashi and the Marton's achievability schemes; meanwhile, it requires a simple rate splitting (3) which is identically exploited also in the Han-Kobayashi scheme for the interference channel. In the full version of the paper [5], we will extend the approach presented here and design a unique achievability scheme for any arbitrary interference network. We actually derive (see [8]) our achievability strategy by a systematic combination of the MAC capacity achieving scheme and the Matron's coding scheme for the BC with common messages (our design in Theorem 1 also falls in this framework). As discussed in introduction, our systematic approach is such that when the designated coding scheme for a given network is specialized to the basic building blocks, it does work essentially similar to the best known coding strategies. Also, each message is split at most into $2^{K-1}$ parts, where $K$ is the number of receivers of the network: Each sub-message is designated to be transmitted to the desired receivers as well as a subset of non-desired receivers. For two-receiver networks each message is split at most into two parts. In fact, for the general networks our achievability design is derived by a simple generalization of the random coding scheme in Fig. 2 on the *MACCM plan of messages*. This plan is described in Section III. Please see also [2].

Let now specialize our achievability scheme for some sub-networks of the BCCR with common message in Fig. 1. Specifically, we consider the BCCR without common message which was previously studied in [19] and [23]. In the following, based on the achievability scheme in Fig. 2, we present a new achievable rate region for this network which may be strictly larger than previous results.

**Corollary:** *Consider the BCCR in Fig. 1 but without common message, i.e., $M_0 \equiv \emptyset$. Define the rate region $\mathfrak{R}_i^{F \to BCCR}$ as (4), where $\mathcal{P}^{BCCR}$ is given by (2). The set $\mathfrak{R}_i^{F \to BCCR}$ constitutes an inner bound on the capacity region.*

*Proof of Corollary)* This achievable rate region is derived by setting $M_0 \equiv \emptyset$ in the achievability scheme of Fig. 2. Let discuss the achievable rate region that is derived by this choice. Consider the rate region (1). When we set $R_0 = 0$, the constraints including $I_{Y_1}^2, I_{Y_2}^2, I_{Y_1}^3, I_{Y_2}^3$ are redundant because they are not corresponding to correctly decoding of any sub-message at its true receiver. In other words, for the case of $M_0 \equiv \emptyset$, without the latter constraints, each receiver can still decode its respective sub-messages with small error probability. The achievable rate region (4) is indeed derived by setting $R_0 = 0$ in (1) and removing redundant constraints. ∎

**Remark:** *By setting $W_B \equiv \emptyset$ and subsequently $B_0 = 0$ in (4), our achievable rate region is directly reduced to that one previously given in [19, Th. 3.2] for the network. This can be verified by a simple comparison. In fact, our rate region may strictly include that of [19, Th. 3.2]. The reason is that our achievable rate region specialized to the two-user BC (see Fig. 3) includes the superposition random variable $W_B$ while this is not the case for the achievability scheme of [19, Th.3.2]. On the one hand, according to [24], for the two-user BC, the Marton's rate region with superposition random variable strictly includes the one without this variable (the region which includes only binning variables, i.e., $U$ and $V$ in Fig. 4). Also, note that the achievable region of [19, Th. 3.2] includes that of [23] as shown in [19, Sec. B]. Thus, our rate region contains previous results.*

### III. EXTENSION TO ARBITRARY SINGLE-HOP TOPOLOGIES

In the full version of this paper [5], we argue that by extending the above approach on the MACCM plan of messages [2], one can derive similar achievability schemes for any other interference network. Below we briefly describe the procedure of construction of the MACCM plan of messages for an arbitrary interference network. Further details will be given in [5].

*MACCM Plan of Messages:* The MACCM plans have a central role in developing our achievability scheme for large multi-user networks. Consider the general interference network as shown in Fig. 5 on the top of the next page.

$$\mathfrak{R}_i^{F \to BCCR} \triangleq \bigcup_{\mathcal{P}^{BCCR}} \left\{ \begin{array}{l} (R_1, R_2) \in \mathbb{R}_+^2 : \exists (B_0, B_1, B_2, R_{10}, R_{11}, R_{20}, R_{22}) \in \mathbb{R}_+^7 : \\ R_1 = R_{10} + R_{11}, \quad R_2 = R_{20} + R_{22}, \\ \\ B_0 > I(U_1, V_2; W_B | W_1, W_2, Q) \triangleq \eta_0 \\ B_0 + B_1 > \eta_0 + I(V_2; U_B | U_1, W_1, W_2, W_B, Q) \triangleq \eta_0 + \eta_1 \\ B_0 + B_2 > \eta_0 + I(U_1; V_B | V_2, W_1, W_2, W_B, Q) \triangleq \eta_0 + \eta_2 \\ B_0 + B_1 + B_2 > \eta_0 + \eta_1 + \eta_2 + I(U_B; V_B | U_1, V_2, W_1, W_2, W_B, Q) \\ \theta_1 \triangleq I(U_1; W_B | W_1, W_2, Q), \\ R_{11} + B_1 < I_{Y_1}^1 + \theta_1 \triangleq I(U_1, U_B; Y_1 | W_1, W_2, W_B, Q) + \theta_1 \\ B_0 + R_{11} + B_1 < I_{Y_1}^4 + \theta_1 \triangleq I(U_1, W_B, U_B; Y_1 | W_1, W_2, Q) + \theta_1 \\ R_{20} + B_0 + R_{11} + B_1 < I_{Y_1}^5 + \theta_1 \triangleq I(U_1, W_2, W_B, U_B; Y_1 | W_1, Q) + \theta_1 \\ R_{10} + B_0 + R_{11} + B_1 < I_{Y_1}^6 + \theta_1 \triangleq I(W_1, U_1, W_B, U_B; Y_1 | W_2, Q) + \theta_1 \\ R_{10} + R_{20} + B_0 + R_{11} + B_1 < I_{Y_1}^7 + \theta_1 \triangleq I(W_1, U_1, W_2, W_B, U_B; Y_1 | Q) + \theta_1 \\ \theta_2 \triangleq I(V_2; W_B | W_1, W_2, Q) \\ R_{22} + B_2 < I_{Y_2}^1 + \theta_2 \triangleq I(V_2, V_B; Y_2 | W_1, W_2, W_B, Q) + \theta_2 \\ B_0 + R_{22} + B_2 < I_{Y_2}^4 + \theta_2 \triangleq I(V_2, W_B, V_B; Y_2 | W_1, W_2, Q) + \theta_2 \\ R_{10} + B_0 + R_{22} + B_2 < I_{Y_2}^5 + \theta_2 \triangleq I(W_1, V_2, W_B, V_B; Y_2 | W_2, Q) + \theta_2 \\ R_{20} + B_0 + R_{22} + B_2 < I_{Y_2}^6 + \theta_2 \triangleq I(W_2, V_2, W_B, V_B; Y_2 | W_1, Q) + \theta_2 \\ R_{10} + R_{20} + B_0 + R_{22} + B_2 < I_{Y_2}^7 + \theta_2 \triangleq I(W_1, V_2, W_2, W_B, V_B; Y_2 | Q) + \theta_2 \end{array} \right\}$$

(4)

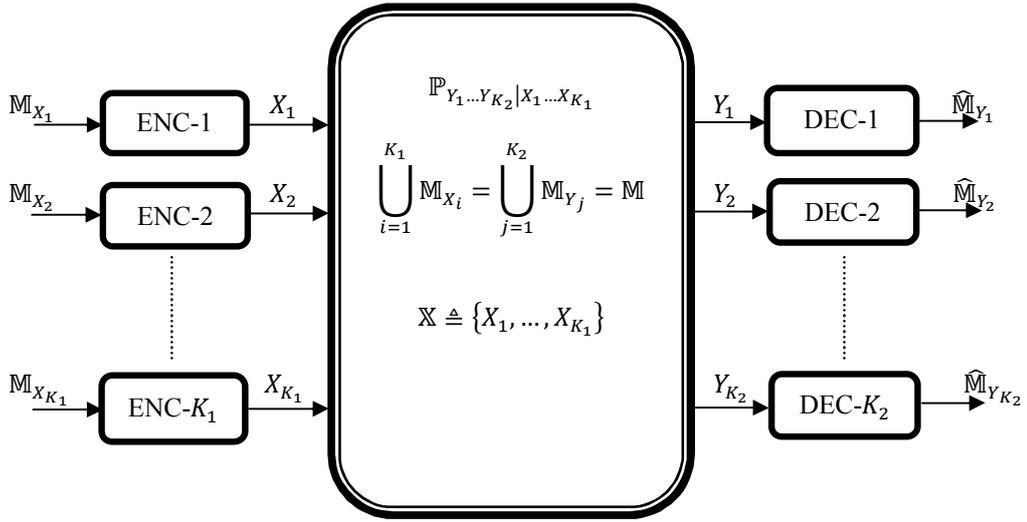

Figure 5. The general interference netwrok.

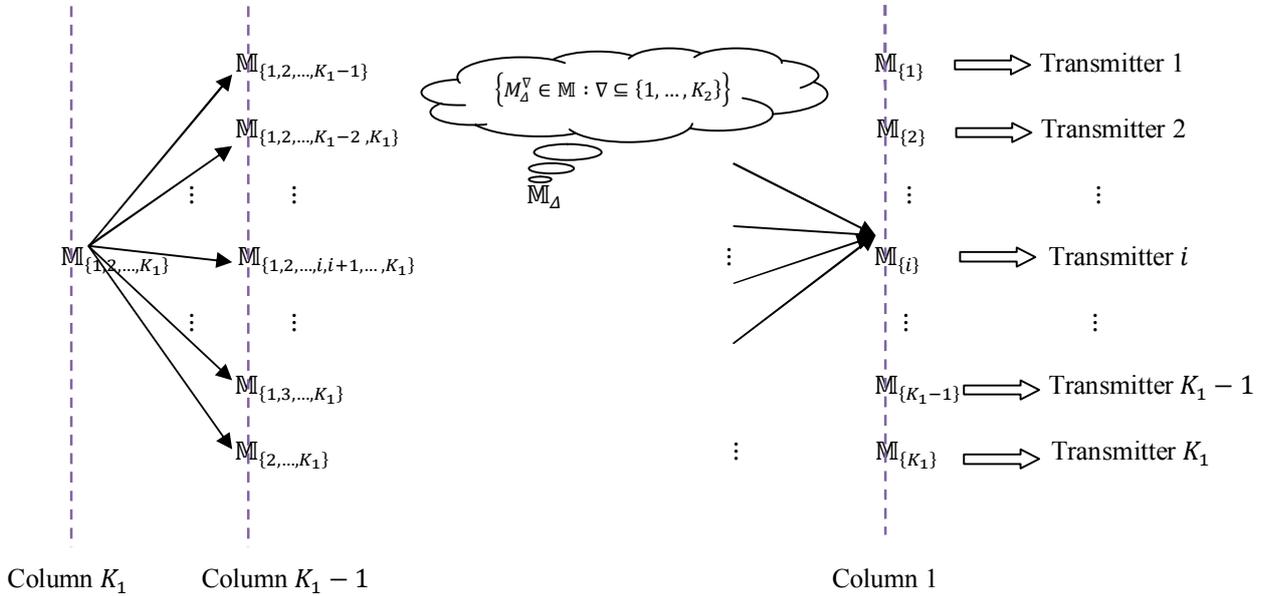

Figure 6. The MACCM plan of messages for an arbitrary interference network.

In this scenario, $K_1$ transmitters send independent messages $\mathbb{M} \triangleq \{M_1, \ldots, M_K\}$ to $K_2$ receivers: the transmitter $X_i$ sends the messages $\mathbb{M}_{X_i}$ over the channel, $i = 1, \ldots, K_1$, and the receiver $Y_j$ decodes the messages $\mathbb{M}_{Y_j}$ for $j = 1, \ldots, K_2$. Therefore, we have:

$$\bigcup_{i=1}^{K_1} \mathbb{M}_{X_i} = \bigcup_{j=1}^{K_2} \mathbb{M}_{Y_j} = \mathbb{M}$$

(5)

The network transition probability function $\mathbb{P}_{Y_1 \ldots Y_{K_2}|X_1 \ldots X_{K_1}}(y_1, \ldots, y_{K_2}|x_1, \ldots, x_{K_1})$ describes the relation between the inputs and the outputsEach subset of transmitters sends at most one message to each subset of receivers. There exist $K_1$ transmitters and $K_2$ receivers. Therefore, we can label each message by a nonempty subset of $\{1, \ldots, K_1\}$ to denote which transmitters send the message, as well as a nonempty subset of $\{1, \ldots, K_2\}$ to determine to which subset of receivers the message is sent. We represent each message of $\mathbb{M}$ as $M_\Delta^\nabla$, where $\Delta \subseteq \{1, \ldots, K_1\}$ and $\nabla \subseteq \{1, \ldots, K_2\}$. For example, $M_{\{1,2,3\}}^{\{2,4\}}$ indicates a message which is sent by Transmitters 1, 2 and 3 to Receivers 2 and 4. Now, for each $\Delta \subseteq \{1, \ldots, K_1\}$, we define:

$$\mathbb{M}_\Delta \triangleq \left\{ M_\Delta^\nabla \in \mathbb{M} : \nabla \subseteq \{1, \ldots, K_2\} \right\}$$

(6)

Using this representation, we arrange the messages into a graph-like illustration as shown in Fig. 6. This illustration is called "MACCM plan of messages". This plan includes $K_1$ columns so that the sets $\mathbb{M}_\Delta, \Delta \subseteq \{1, \ldots, K_1\}$ with $\|\Delta\| = i$ are situated in its $i^{th}$ column, $i = 1, \ldots, K_1$. Also, the set $\mathbb{M}_{\Delta_1}$ in column $i, i = 2, \ldots, K_1$, is connected (by a directed edge) to the set $\mathbb{M}_{\Delta_2}$ in column $i - 1$ provided that $\Delta_2 \subseteq \Delta_1$. Please refer to [2] for details.

If the network includes only one receiver, then it is reduced to a MAC with common messages. In this case, each of the sets $\mathbb{M}_\Delta$ contains at most one message. Thus, the MACCM plan of messages coincides to the MACCM message graph [2] representing the superposition coding scheme which achieves the capacity region for the channel. In this scheme, the codewords corresponding to every two messages connected by a directed edge build a superposition structure such that the message at the beginning of the edge is the cloud center and that at the end of the edge is the satellite.

According to the MACCM plan, the messages $\mathbb{M}_\Delta$ given in (6) are broadcasted by the transmitters $X_i, i \in \Delta$, meanwhile, no transmitter other than those in $\{X_i, i \in \Delta\}$ has access to these messages. To derive our achievability scheme for the general interference networks, first an efficient message splitting strategy is chosen. Then, considering the modified network which is derived by the message splitting scheme, the messages belonging to each of the sets $\mathbb{M}_\Delta$ are encoded according to a suitable broadcasting strategy. For example, for the two-receiver networks, a Marton's type encoding is built to encode these messages (note that for two-receiver networks each of the message sets $\mathbb{M}_\Delta$ includes at most three messages). The achievability scheme in Fig. 2 for the BCCR with common message is indeed an expressive example for our approach. The details are reported in the full version of the paper.

CONCLUSION

The purpose of this paper was to show that for any given single-hop communication network with two receivers, splitting messages into more than two sub-messages in a random coding scheme is redundant. To this end, we considered the BCCR with common message. We presented a novel achievability scheme for this network. Our achievability design is derived by a systematic combination of the best known achievability schemes for the basic building blocks included in the network: the Han-Kobayashi scheme for the two-user classical interference channel and the Marton's coding scheme for the broadcast channel. Meanwhile, in our scheme each private message is split into only two sub-messages which is identically exploited also in the Han-Kobayashi scheme. We justified that our achievable rate region strictly includes also previous results. More importantly, we provided a graphical illustration for the achievability scheme based on the directed graphs which were developed in our previous paper [7, 8] to describe/design random coding schemes. Then, we argued that by extending the proposed scheme on the MACCM plan of messages, one can derive similar achievability schemes for any other interference network.

ACKNOWLEDGEMENT

The author would like to thank F. Marvasti whose editing comments improved the language of this paper.

REFERENCES

[1] R. K. Farsani, "Fundamental limits of communications in interference networks-Part I: Basic structures," *IEEE Trans. Information Theory, Submitted for Publication, 2012, available at* http://arxiv.org/abs/1207.3018.

[2] __________, "Fundamental limits of communications in interference networks-Part II: Information flow in degraded networks," *IEEE Trans. Information Theory, Submitted for Publication, 2012, available at* http://arxiv.org/abs/1207.3027.

[3] __________, "Fundamental limits of communications in interference networks-Part III: Information flow in strong interference regime," *IEEE Trans. Information Theory, Submitted for Publication, 2012, available at* http://arxiv.org/abs/1207.3035.

[4] __________, "Fundamental limits of communications in interference networks-Part IV: Networks with a sequence of less-noisy receivers," *IEEE Trans. Information Theory, Submitted for Publication, 2012, available at* http://arxiv.org/abs/1207.3040.

[5] __________, "Fundamental limits of communications in interference networks-Part V: A random coding scheme for transmission of general message sets," *IEEE Trans. Information Theory, To be submitted, preprint available at* http://arxiv.org/abs/1107.1839.

[6] __________, " Capacity theorems for the cognitive radio channel with confidential messages," 2012, available at http://arxiv.org/abs/1207.5040.

[7] R. K. Farsani, F. Marvasti, "Interference networks with general message sets: A random coding scheme" [Online] *available at*: http://arxiv.org/abs/1107.1839.

[8] __________, "Interference networks with general message sets: A random coding scheme", 2011, 49th *Annual Allerton Conference on Communication, Control, and Computing.*, Monticello, IL, Sep. 2011.

[9] A. El Gamal and Y.-H. Kim, *Lecture notes on network information theory*, arXiv:1001.3404, 2010.

[10] N. Devroye, P. Mitran, and V. Tarokh, "Achievable rates in cognitive radio channels," *IEEE Trans. Inf. Theory*, vol. 52, pp. 1813–1827, May 2006.

[11] A. Jovicic and P. Viswanath, "Cognitive radio: An information-theoretic perspective," *IEEE Int. Symp. Inf. Theory*, Jul. 2006, pp. 2413–2417.

[12] I. Maric, R. Yates, and G. Kramer, "Capacity of interference channels with partial transmitter cooperation," *IEEE Trans. Inf. Theory*, vol. 53, no. 10, pp. 3536–3548, Oct. 2007.

[13] S. Rini, D. Tuninetti, and N. Devroye, "State of the cognitive interference channel: A new unified inner bound," in *Int. Zurich Seminar on Communications (IZS)*, Mar. 2010, pp. 57–62.

[14] Y. Cao and B. Chen, "Interference channels with one cognitive transmitter," *Asilomar Conference on Signals, Systems, and Computers*, Nov. 2008 Arxiv preprint arXiv:0910.0899.

[15] W. Wu, S. Vishwanath, and A. Arapostathis, "Capacity of a class of cognitive radio channels: Interference channels with degraded message sets," *IEEE Trans. Inf. Theory*, vol. 53, no. 11, pp. 4391–4399, Nov. 2007.

[16] J. Jiang and Y. Xin, "On the achievable rate regions for interference channels with degraded message sets," *IEEE Trans. Inf. Theory*, vol. 54, no. 10, pp. 4707–4712, 2008.

[17] I. Maric, A. Goldsmith, G. Kramer, and S. Shamai, "On the capacity of interference channels with one cooperating transmitter," *European Trans. Telecomm.*, vol. 19, pp. 405–420, Apr. 2008.

[18] G. Hodtani and M. Aref, "On the Devroye-Mitran-Tarokh rate region for the cognitive radio channel," *IEEE Trans. Wireless Comm.*, vol. 8, no. 7, pp. 3458–3461, 2009.

[19] J. Jiang, I. Maric, A. Goldsmith, and S. Cui, "Achievable rate regions for broadcast channels with cognitive relays," in *Proc. IEEE Information Theory Workshop (ITW)*, Taormina, Italy, 2009, pp. 500–504.

[20] N. Liu, I. Maric, A. Goldsmith, and S. Shamai, "Capacity bounds and exact results for the cognitive Z-interference channel," available at arXiv:1112.2483.

[21] J. Jiang, I. Maric, A. Goldsmith, S. Shamai, and S. Cui, "On the capacity of a class of cognitive Z-interference channels," In *Proc IEEE Int. Conference on Communications (ICC)*, Kyoto, Japan, pp.1-6, Jun. 2011.

[22] M. Vaezi and M. Vu, "On the capacity of the cognitive Z-interference channel," *Canadian Workshop on Information Theory (CWIT)*, Kelowna, BC, pp. 30-33, May 2011.

[23] S. Sridharan, S. Vishwanath, S. Jafar, and S. Shamai, "On the capacity of cognitive relay assisted Gaussian interference channel," in *Proc. IEEE Int. Symp. Inf. Theory*. Toronto: IEEE, 2008, pp. 549–553.

[24] A. Gohari, A. El Gamal, and V. Anantharam, "On Marton's inner bound for the general broadcast channel," *IEEE Trans. Inf. Theory, Submitted for Publication*. arXiv:1006.5166.